# Measuring Feature-Label Dependence Using Projection Correlation Statistic

Yixiao Liu[a]* and Pengjian Shang[b]

[a]*School of Mathematics and Statistics, Beijing Jiaotong University, Beijing 100044, PR China*

[b]*School of Mathematics and Statistics, Beijing Jiaotong University, Beijing 100044, PR China*

*Corresponding author

E-mail address: 23121729@bjtu.edu.cn (Yixiao Liu); pjshang@bjtu.edu.cn (Pengjian Shang)

# Measuring Feature-Label Dependence Using Projection Correlation Statistic

## Abstract

Detecting dependence between variables is a crucial issue in statistical science. In this paper, we propose a novel metric, named label projection correlation, to measure the dependence between numerical and categorical variables. The proposed correlation does not require any conditions on the numerical variable, and is equal to zero if and only if the two variables are independent. Moreover, when the numerical variable is one-dimensional, we demonstrate that the computational cost of the correlation estimation can be reduced from $O(n^3)$ to $O(n\log n)$, where $n$ is the sample size. Furthermore, if the one-dimensional variable is continuous, the metric can be simplified to a concise rank-based expression. The asymptotic theorem of the estimation is also established. Two simulated experiments are presented to demonstrate the effectiveness of our proposed correlation in feature selection. Furthermore, our approach is applied to feature selection in drivers' facial images and cancer mass-spectrometric data.

**Keywords**: Independence measure, Projection correlation, Categorical variable, Feature selection

# 1. Introduction

It is fundamental to quantify the dependence between random variables in statistical research. Since the introduction of the Pearson correlation coefficient, several metrics have been proposed to capture the dependence between two random variables, such as Spearman's and Kendall's correlations, mutual information (Duncan 1970) and the maximal information coefficient (Reshef et al. 2011).

Among these metrics, a special type of dependence measure arises from measuring independence. A representative example is the distance correlation (Székely et al. 2007), which is zero if and only if two random variables are independent. Sejdinovic et al. (2013) further established the equivalence of distance correlation and another dependence measure, the Hilbert-Schmidt Independence Criterion (HSIC) (Gretton et al. 2005). Based on the distance correlation, Wang et al. (2015) constructed the conditional distance correlation, which measures the conditional dependence. Furthermore, Zhu et al. (2017) proposed the projection correlation, which overcomes the moment limitation of random variables. Kong et al. (2019) developed a composite coefficient of determination, and Pan et al. (2020) developed ball covariance to explore dependence in Banach space. Chatterjee (2021) constructed a new rank-based coefficient that detects functional dependence between two random variables, and the coefficient has been extended to measure conditional dependence (Azadkia and Chatterjee 2021). In addition, certain metrics are designed to capture specific forms of dependence. A notable example is the martingale difference correlation (MDC) proposed by Shao and Zhang (2014), which tests mean dependence between two variables, along with several subsequent refinements (Li et al. 2023; Zhou and Zhu 2021).

These methods are mostly used to measure the correlation between two numerical variables. However, metrics used to quantify the relationship between numerical and categorical variables are also considered. Cui et al. (2015) proposed the MV method based on the Cramér-von Mises distance, Dang et al. (2021) constructed the Gini distance correlation, and Zhang et al. (2021) proposed the Reproducing Kernel Hilbert Space (RKHS) based Gini distance correlation. However, it is worth noting that each of these methods comes with its own set of limitations. The MV method can only be applied to measure the dependence between one-dimensional numerical variable and categorical variable. Gini distance correlation requires moment limitations, and RKHS-based Gini distance correlation depends on the selection of Mercer Kernel and parameter setting.

Therefore, it is essential to develop a novel dependence measure that not only requires fewer limitations on random variables, but also is free of choosing kernel and parameter. In this paper, we propose the label projection correlation to detect the dependence between numerical and categorical variables. The correlation does not require limitations on numerical variables, equals zero if and only if the two variables are independent, and is invariant to orthogonal transformation. The computational cost of its estimation is $O(n^3)$, where $n$ is the sample size. However, when the numerical variable is univariate, we demonstrate that the computational cost of its estimation can be reduced to $O(n\log n)$. Moreover, when the numerical variable is also continuous, the correlation can be represented as a concise expression with the rank of each sample in the full sample and its categorical sample. Furthermore, the asymptotic properties of the estimation are established. Specifically, for independent variables, the estimation converges in distribution to a weighted sum of chi-square random variables, and for dependent variables, it demonstrates asymptotic normality. Two simulation experiments demonstrate the effectiveness of our method in feature selection, and this approach is further applied to two real data analyses.

The rest of this article is organized as follows. In section 2, we show the definition of label projection correlation, provide its statistical properties and estimation, and establish the asymptotic theorems of its estimation. In section 3, numerical experiments are conducted to demonstrate the performance of our method in feature selection. Finally, the conclusions are summarized in section 4. Technical proofs are presented in the Appendix.

# 2. Label Projection Correlation

### *2.1 Motivation*

It is fundamental to investigate the dependence between categorical and numerical variables. Cui et al. (2015) proposed the MV method based on the Cramér-von Mises distance and gave its feature screening procedure. However, this method can only measure the relationship between one-dimensional numerical variable and categorical variable. To address this limitation, Dang et al. (2021) have proposed the Gini distance correlation, which could be expressed as

$$\mathrm{gCor}(X,Y)=\frac{E\left\|X_1-X_2\right\|-\sum_{k=1}^{K}p_kE\left\|X_1^{(k)}-X_2^{(k)}\right\|}{E\left\|X_1-X_2\right\|}. \tag{1}$$

In Eq.(1), $X\in\mathbb{R}^p$ is the numerical variable, $Y\in\mathbb{Y}=\{1,\ldots,K\}$ is the categorical variable with $K$ different labels and $P(Y=k)=p_k$. Under $Y=k$, the numerical variable is denoted as $X^{(k)}$. $\{X_1,X_2\}$ and $\{X_1^{(k)},X_2^{(k)}\}$ are the independent copies of $X$ and $X^{(k)}$, respectively. In the following discussion, we assume that the fixed categorical number $K\geq 2$, and $p_i>0$ for all $i\in\{1,\ldots,K\}$.

However, there is an undeniable limitation of Gini distance correlation that $X$ must satisfies $E\left\|X\right\|<+\infty$. To address it, Zhang et al.(2021) considered the Gini distance correlation in RKHS, which can be expressed as:

$$gKCor(X,Y)=\frac{Ed\left(X_1,X_2\right)-\sum_{k=1}^{K}p_kEd\left(X_1^{(k)},X_2^{(k)}\right)}{Ed\left(X_1,X_2\right)}, \tag{2}$$

where $d(x,y)=\sqrt{\kappa(x,x)+\kappa(y,y)-2\kappa(x,y)}$ and $\kappa(x,y)$ is the Mercer Kernel. This measure overcomes the moment limitation in Gini distance correlation. However, the selection and parameter settings of the Mercer Kernel become new challenges. To avoid these issues, we adopt the thought of projection correlation (Zhu et al. 2017) to construct a novel dependence measure without conditions on the numerical variable.

## *2.2 Label Projection Correlation*

Testing $X \in \mathbb{R}^p$ and $Y \in \mathbb{Y}$ are independent is equivalent to testing whether $U=\boldsymbol{\alpha}^T X$ and $Y$ are independent for all unit vectors $\boldsymbol{\alpha} \in \mathbb{R}^p$. Let $F_U(u)$ and $F_{U|Y=k}(u)$ denote the cumulative distribution function (c.d.f) of $U$ and the conditional c.d.f of $U$ under $Y=k$, respectively. Given $\boldsymbol{\alpha}$, $U$ and $Y$ are independent if and only if $F_{U|Y=k}(u)=F_U(u)$ for all $u \in S$, where $S$ is the support set of $U$. Therefore, we construct the following metric:

$$\sum_{k=1}^{K} p_k \int_{\|\boldsymbol{\alpha}\|=1}\int_{\mathbb{R}} (F_{U|Y=k}(u)-F_U(u))^2\, dF_U(u)d\boldsymbol{\alpha}, \tag{3}$$

which is equal to zero if and only if $X$ and $Y$ are independent, and it can also be expressed in the form of indicator function:

$$\begin{aligned}
&\sum_{k=1}^{K} p_k \int_{\|\boldsymbol{\alpha}\|=1}\int_{\mathbb{R}} (E(I(U\le u)\mid Y=k) - E(I(U\le u)))^2\, dF_U(u)d\boldsymbol{\alpha} \\
&=\sum_{k=1}^{K} \frac{1}{p_k} \int_{\|\boldsymbol{\alpha}\|=1}\int_{\mathbb{R}} (E(I(U\le u)I(Y=k)) - E(I(U\le u))E(I(Y=k)))^2\, dF_U(u)d\boldsymbol{\alpha}
\end{aligned}. \tag{4}$$

Utilizing the Fubini theorem, Eq.(4) can be expressed as follows:

$$\begin{aligned}
&\sum_{k=1}^{K}\frac{1}{p_k}E[\int_{\|\boldsymbol{\alpha}\|=1}\int_{\mathbb{R}}\left(I(U_1\le u)I(Y_1=k)-I(U_1\le u)I(Y_2=k)\right)\left(I(U_3\le u)I(Y_3=k)-I(U_3\le u)I(Y_4=k)\right)dF_U(u)d\boldsymbol{\alpha}] \\
&=\sum_{k=1}^{K}\frac{1}{p_k}E[\int_{\|\boldsymbol{\alpha}\|=1} I(U_1\le U_5)I(U_3\le U_5)I(Y_1=k)I(Y_3=k) - 2I(U_1\le U_5)I(U_3\le U_5)I(Y_2=k)I(Y_3=k) \\
&\quad + I(U_1\le U_5)I(U_3\le U_5)I(Y_2=k)I(Y_4=k)d\boldsymbol{\alpha}] \\
&=\sum_{k=1}^{K}\frac{1}{p_k}E[\int_{\|\boldsymbol{\alpha}\|=1} I(U_1\le U_5)I(U_3\le U_5)I(Y_1=k)I(Y_3=k)\mathrm{d}\boldsymbol{\alpha}] - E[\int_{\|\boldsymbol{\alpha}\|=1} I(U_1\le U_5)I(U_3\le U_5)\mathrm{d}\boldsymbol{\alpha}] \\
&=\sum_{k=1}^{K}\frac{1}{p_k}E[\int_{\|\boldsymbol{\alpha}\|=1} I(\boldsymbol{\alpha}^T(X_1-X_3)\le 0)I(\boldsymbol{\alpha}^T(X_2-X_3)\le 0)I(Y_1=k)I(Y_2=k)\mathrm{d}\boldsymbol{\alpha}] - E[\int_{\|\boldsymbol{\alpha}\|=1} I(\boldsymbol{\alpha}^T(X_1-X_3)\le 0)I(\boldsymbol{\alpha}^T(X_2-X_3)\le 0)\mathrm{d}\boldsymbol{\alpha}], \\
&=\sum_{k=1}^{K}p_k E[\int_{\|\boldsymbol{\alpha}\|=1} I(\boldsymbol{\alpha}^T(X_1-X_3)\le 0)I(\boldsymbol{\alpha}^T(X_2-X_3)\le 0)\mathrm{d}\boldsymbol{\alpha} \mid Y_1=k, Y_2=k] - E[\int_{\|\boldsymbol{\alpha}\|=1} I(\boldsymbol{\alpha}^T(X_1-X_3)\le 0)I(\boldsymbol{\alpha}^T(X_2-X_3)\le 0)\mathrm{d}\boldsymbol{\alpha}],
\end{aligned} \tag{5}$$

where $\{(U_i, X_i, Y_i), i=1,2,3,4,5\}$ are the independent copies of $(U,X,Y)$. However, the final expression in Eq.(5) is quite complex. Lemma 1 enables us to simplify it.

***Lemma 1*** *(Escanciano 2006) For two arbitrary non-zero vectors* $U_1, U_2 \in \mathbb{R}^p$*, we have*

$$\int_{\|\boldsymbol{\alpha}\|=1} I(\boldsymbol{\alpha}^{\mathrm{T}}U_1 \leqslant 0)I(\boldsymbol{\alpha}^{\mathrm{T}}U_2 \leqslant 0)\,\mathrm{d}\boldsymbol{\alpha} = c_p\left\{\pi - \arccos\left(\frac{U_1^{\mathrm{T}}U_2}{\|U_1\|\|U_2\|}\right)\right\},$$

*where* $c_p = \pi^{p/2-1}/\Gamma(p/2)$ *and* $\Gamma(\bullet)$ *is the gamma function.* $c_p$ *is the surface area of a p-dimensional unit ball divided by* $2\pi$*.*

For other cases, it is natural that

$$\int_{\|\boldsymbol{\alpha}\|=1} I(\boldsymbol{\alpha}^{\mathrm{T}}U_1 \leqslant 0)I(\boldsymbol{\alpha}^{\mathrm{T}}U_2 \leqslant 0)\,\mathrm{d}\boldsymbol{\alpha} = \begin{cases} \pi c_p, & U_i=\mathbf{0}, U_j \ne \mathbf{0}\ (i\ne j); \\ 2\pi c_p, & U_1=U_2=\mathbf{0}. \end{cases}$$

For the simplicity of notation, in the following discussion, we still define the two special cases in the form of $\arccos(\bullet)$, which indicates

$$\arccos\left(\frac{U_1^{\mathrm{T}} U_2}{\|U_1\|\|U_2\|}\right)=\begin{cases}0, U_i=\mathbf{0}, U_j \neq \mathbf{0}\,(i \neq j); \\ -\pi, U_1=U_2=\mathbf{0}.\end{cases}$$

According to Lemma 1, Eq.(5) could be simplified as

$$c_p\left(E(\arccos(\frac{(X_1-X_3)^T(X_2-X_3)}{\|X_1-X_3\|\|X_2-X_3\|}))-\sum_{k=1}^K p_k E(\arccos(\frac{(X_1-X_3)^T(X_2-X_3)}{\|X_1-X_3\|\|X_2-X_3\|}) \mid Y_1=k, Y_2=k)\right). \tag{6}$$

To express it concisely, we define

$$\begin{aligned} & LPCov(X, Y)=S_1-S_2 \\ & =E(\arccos(\frac{(X_1-X_3)^T(X_2-X_3)}{\|X_1-X_3\|\|X_2-X_3\|}))-\sum_{k=1}^K p_k E(\arccos(\frac{(X_1-X_3)^T(X_2-X_3)}{\|X_1-X_3\|\|X_2-X_3\|}) \mid Y_1=k, Y_2=k),\end{aligned} \tag{7}$$

where $S_1$ and $S_2$ are defined in the obvious manner. Furthermore, Liu et.al (2022) and Wang et.al(2025) have proposed independence testing methods based on this measure. Then, we consider the normalization of it. First, $LPCov(X,Y)$ can be expressed as:

$$\begin{aligned} LPCov(X, Y) & =\frac{1}{c_p} \sum_{k=1}^K p_k \int_{\|\boldsymbol{\alpha}\|=1} \int_{\mathbb{R}}(F_{\boldsymbol{\alpha}^T X \mid Y=k}(u)-F_{\boldsymbol{\alpha}^T X}(u))^2 d F_{\boldsymbol{\alpha}^T X}(u) d \boldsymbol{\alpha} \\ & =\frac{1}{c_p} \int_{\|\boldsymbol{\alpha}\|=1} E_{X_1}\left[Var_Y\left[E_X\left[I\left(\boldsymbol{\alpha}^T X \leq \boldsymbol{\alpha}^T X_1\right) \mid Y\right]\right]\right] d \boldsymbol{\alpha}.\end{aligned} \tag{8}$$

The law of total variance shown in Lemma 2 can be applied to normalize it.

***Lemma 2***. *Law of Total Variance*

$$Var(X)=Var(E(X \mid Y))+E(Var(X \mid Y)).$$

According to Lemma 2, the upper boundary of $LPCov(X,Y)$ is shown below:

$$\begin{aligned} LPCov_{upper}(X, Y) & =\frac{1}{c_p} \int_{\|\boldsymbol{\alpha}\|=1} E_{X_1}\left[Var_X\left(I\left(\boldsymbol{\alpha}^T X \leq \boldsymbol{\alpha}^T X_1\right)\right)\right] d \boldsymbol{\alpha} \\ & =\frac{1}{c_p} \int_{\|\boldsymbol{\alpha}\|=1} E\left[I\left(\boldsymbol{\alpha}^T X_2 \leq \boldsymbol{\alpha}^T X_1\right)\left(1-I\left(\boldsymbol{\alpha}^T X_3 \leq \boldsymbol{\alpha}^T X_1\right)\right)\right] d \boldsymbol{\alpha} \\ & =\frac{1}{c_p} \int_{\|\boldsymbol{\alpha}\|=1} E\left[I\left(\boldsymbol{\alpha}^T X_2 \leq \boldsymbol{\alpha}^T X_1\right)\right]-E\left[I\left(\boldsymbol{\alpha}^T X_2 \leq \boldsymbol{\alpha}^T X_1\right) I\left(\boldsymbol{\alpha}^T X_3 \leq \boldsymbol{\alpha}^T X_1\right)\right] d \boldsymbol{\alpha} \\ & =\left(\pi+\pi E\left[I\left(X_1=X_2\right)\right]\right)-\left(\pi-E\left[\arccos(\frac{(X_1-X_3)^T(X_2-X_3)}{\|X_1-X_3\|\|X_2-X_3\|})\right]\right) \\ & =\pi E\left[I\left(X_1=X_2\right)\right]+E\left[\arccos(\frac{(X_1-X_3)^T(X_2-X_3)}{\|X_1-X_3\|\|X_2-X_3\|})\right] \\ & =\pi S_3+S_1,\end{aligned} \tag{9}$$

where $S_3=E[I(X_1=X_2)]$. We define the normalized $LPCov(X,Y)$ as the label projection correlation $LPC(X,Y)$ shown in Eq.(10):

$$LPC(X, Y)=\frac{S_1-S_2}{\pi S_3+S_1}. \tag{10}$$

We set $LPC(X,Y)=0$ when $\pi S_3+S_1=0$ (i.e., $X$ is a constant vector almost surely). The constructed measure overcomes the moment limitation in the Gini distance correlation since it is an angle-based measure. Theorem 1 summarizes some statistical properties of label projection correlation, and their proofs are given in the Appendix.

***Theorem 1.*** *For numerical variable* $X \in \mathbb{R}^p$ *and categorical variable* $Y \in \mathbb{Y}$ with fixed categorical number $K$, $LPC(X,Y)$ *has the following properties.*

(1) $0 \le LPC(X,Y) \le 1$.

(2) $LPC(X,Y) = 0$ *if and only if* $X$ *and* $Y$ *are independent.*

(3) $LPC(X,Y) = 1$ *if and only if for all label* $k$, $X^{(k)}$ *is equal to a constant vector* $\boldsymbol{c_k}$, *and these* $\boldsymbol{c_k}$ *are not all the same.*

(4) $LPC(X,Y) = LPC(a\boldsymbol{C}X + \boldsymbol{b}, Y)$ *for any orthogonal matrix* $\boldsymbol{C} \in \mathbb{R}^{p \times p}$, *nonzero constant* $a$ *and arbitrary vector* $\boldsymbol{b} \in \mathbb{R}^p$.

(5) *When* $X$ *is a continuous random variable,* $LPC(X,Y) = \dfrac{\pi - 3S_2}{\pi}$.

(6) *Suppose that* $X^{(k)} \sim N(\boldsymbol{0}, \sigma_k^2 \boldsymbol{I}_p)$, *where* $\boldsymbol{I}_p$ *is the p-dimensional identity matrix, then*

$$LPC(X,Y) = \frac{3}{\pi}\sum_{i=1}^{K}\sum_{j=1}^{K} p_i p_j \left[\frac{1}{2}\arcsin\left(\frac{\sigma_i^2}{\sigma_i^2 + \sigma_j^2}\right) + \frac{1}{2}\arcsin\left(\frac{\sigma_j^2}{\sigma_i^2 + \sigma_j^2}\right) - \frac{\pi}{6}\right].$$

A similar expression and several properties of $LPC(X,Y)$ have also appeared in the literature (Wang et al. 2025), but our metric has been further refined to accommodate cases where numerical variable $X$ is non-continuous, thereby endowing it with greater generalizability.

## *2.3 Estimation*

The estimation of $LPC(X,Y)$ is given in this section. We first denote

$$a_{ijl} = \arccos\left(\frac{(X_i - X_l)^T (X_j - X_l)}{\|X_i - X_l\|\|X_j - X_l\|}\right), \tag{11}$$

let $a_{ijl} = 0$ when only one of $\|X_i - X_l\|$ and $\|X_j - X_l\|$ is zero, $a_{ijl} = -\pi$ when both $\|X_i - X_l\|$ and $\|X_j - X_l\|$ are zero. The estimation of $S_1$, $S_2$ and $S_3$ could be expressed as:

$$\hat{S}_1 = \frac{1}{n^3}\sum_{1 \le i,j,l \le n} a_{ijl}, \tag{12}$$

$$\hat{S}_2 = \sum_{k=1}^{K} \hat{p}_k \frac{1}{nn_k^2}\sum_{\substack{i,j \in \Theta_k \\ 1 \le l \le n}} a_{ijl} = \frac{1}{n^2}\sum_{k=1}^{K}\frac{1}{n_k}\sum_{\substack{i,j \in \Theta_k \\ 1 \le l \le n}} a_{ijl}, \tag{13}$$

$$\hat{S}_3 = \frac{1}{n^2}\sum_{j=1}^{n}\sum_{i=1}^{n} I(X_i = X_j), \tag{14}$$

where $\Theta_k$ refers to all samples with label $Y = k$, $n_k$ is the sample size of $\Theta_k$, and $\hat{p}_k = n_k / n$. $LPC(X,Y)$ can be estimated as follows:

$$\widehat{LPC}_n(X,Y) = \frac{\hat{S}_1 - \hat{S}_2}{\hat{S}_1 + \pi\hat{S}_3}, \tag{15}$$

and the computational cost of it is $O(n^3)$. The following theorem states the strong consistency of it, and its proof is presented in the appendix.

***Theorem 2***. *For the given random sample* $\{X_i, Y_i\}_{i=1}^n$, $\widehat{LPC}_n(X,Y)$ *converges to* $LPC(X,Y)$ *almost surely.*

## *2.4 One-dimensional Label Projection Correlation*

In this section, we present a simplified expression for $LPC(X,Y)$ and a fast algorithm for its estimation in the case where $X$ is univariate. When $X$ is univariate, the projection vector $\boldsymbol{\alpha}$ degenerates to -1 and 1. Therefore, $LPC(X,Y)$ can be expressed as follows:

$$LPC(X,Y)=\frac{E_{X_1}\left[Var_Y\left[E_X\left[I\left(X\leq X_1\right)|Y\right]\right]\right]+E_{X_1}\left[Var_Y\left[E_X\left[I\left(X\geq X_1\right)|Y\right]\right]\right]}{E_{X_1}\left[Var_X\left(I\left(X\leq X_1\right)\right)\right]+E_{X_1}\left[Var_X\left(I\left(X\geq X_1\right)\right)\right]}, \tag{16}$$

Lemma 3 enables us to simplify it, and its proof is provided in the Appendix.

***Lemma 3.*** *Denote* $Q$ *as* $F_X(X)$, $\{Q_1,Q_2\}$ *as the independent copies of* $Q$, $Q^{(k)}$ *as* $F_X(X^{(k)})$, $\{Q_1^{(k)},Q_2^{(k)}\}$ *as the independent copies of* $Q^{(k)}$. *We have*

$$\begin{aligned}&E(|Q_1-Q_2|)=2E_{X_2}\left(Var\left(I\left(X_1\geq X_2\right)\right)\right),\\&\sum_{k=1}p_kE\left(\left|Q_1^{(k)}-Q_2^{(k)}\right|\right)=2E_{X_2}\left(E_{Y_1}\left(Var_{X_1}\left(I\left(X_1\geq X_2\right)|Y_1\right)\right)\right).\end{aligned} \tag{17}$$

According to Lemma 2 and Lemma 3, it is obvious that

$$E_{X_1}\left[Var_Y\left[E_X\left[I\left(X\geq X_1\right)|Y\right]\right]\right]=\frac{1}{2}\left(E\left(|Q_1-Q_2|\right)-\sum_{k=1}p_kE\left(\left|Q_1^{(k)}-Q_2^{(k)}\right|\right)\right). \tag{18}$$

Following the definition of $Q$, $\{Q_1,Q_2\}$, $Q^{(k)}$, $\{Q_1^{(k)},Q_2^{(k)}\}$, the variables $^{-}Q$, $\{^{-}Q_1,{}^{-}Q_2\}$, $^{-}Q^{(k)}$, $\{^{-}Q_1^{(k)},{}^{-}Q_2^{(k)}\}$ are defined as the same manner of $-X$. For simplicity, we define $T_1=E\left(\left|Q_1-Q_2\right|\right)$, $T_2=\sum_{k=1}^{K}p_kE\left(\left|Q_1^{(k)}-Q_2^{(k)}\right|\right)$, $^{-}T_1=E\left(\left|{}^{-}Q_1-{}^{-}Q_2\right|\right)$, and $^{-}T_2=\sum_{k=1}^{K}p_kE\left(\left|{}^{-}Q_1^{(k)}-{}^{-}Q_2^{(k)}\right|\right)$. Hence Eq.(16) can be expressed as

$$LPC(X,Y)=\frac{\left(T_1+{}^{-}T_1\right)-\left(T_2+{}^{-}T_2\right)}{T_1+{}^{-}T_1}. \tag{19}$$

We provide a fast algorithm for estimating it. Here, we denote $Q_i=\hat{F}_X(X_i)=\sum_{j=1}^{n}I(X_j\leq X_i)\big/n$, $Q_i^{(k)}=\hat{F}_X(X_i^{(k)})=\sum_{j=1}^{n}I(X_j\leq X_i,i\in\Theta_k)\big/n$, and arrange them in ascending order as $\{Q_{(i)}\}_{i=1}^{n}$ and $\{Q_{(i)}^{(k)}\}_{i=1}^{n_k}$, respectively. $T_1$ and $T_2$ are estimated as

$$\hat{T}_1=\frac{2}{n^2}\sum_{i>j}^{n}\left|Q_i-Q_j\right|=\frac{2}{n^2}\sum_{i>j}^{n}\left(Q_{(i)}-Q_{(j)}\right)=\frac{2}{n^2}\sum_{i=1}^{n}\left(2i-n-1\right)Q_{(i)}, \tag{20}$$

$$\hat{T}_2=\sum_{k=1}^{K}\hat{p}_k\frac{2}{n_k^2}\sum_{i>j}^{n_k}\left|Q_i^{(k)}-Q_j^{(k)}\right|=\sum_{k=1}^{K}\frac{2\hat{p}_k}{n_k^2}\sum_{i=1}^{n_k}\left(2i-n_k-1\right)Q_{(i)}^{(k)}=\frac{2}{n}\sum_{k=1}^{K}\frac{1}{n_k}\sum_{i=1}^{n_k}\left(2i-n_k-1\right)Q_{(i)}^{(k)}. \tag{21}$$

$^{-}T_1$ and $^{-}T_2$ can be estimated in the same manner. The estimator derived from these statistics by Eq.(19) is consistent with the estimator shown in Eq.(15). The estimating algorithm only requires the rank of each sample in the full sample and its categorical sample. As the computational cost of sorting algorithm can be reduced to $O(n\log n)$, the computational cost of $LPC(X,Y)$ can be decreased to $O(n\log n)$ in this way.

Notably, Theorem 3 shows a more concise rank-based expression and estimator of $LPC(X,Y)$ when $X$ is continuous, and the proof of it is given in the Appendix.

***Theorem 3***. *When $X$ is a one-dimensional continuous random variable and $Y$ has a fixed categorical number $K$,*

(1) *$LPC(X,Y)$ can be simplified as follows*:

$$LPC(X,Y) = 4(1 - 3\sum_{k=1}^{K} p_k E(Q^{(k)} F_{X^{(k)}}(X^{(k)}))). \tag{22}$$

(2) *$LPC(X,Y)$ can be estimated according to Eq.(22), and this estimator is expressed as*:

$$\widehat{LPC}_n(X,Y) = 4 - 12\sum_{k=1}^{K}\frac{n_k}{n}\frac{1}{n_k}\sum_{i=1}^{n_k}\frac{i}{n_k}Q_{(i)}^{(k)} = 4 - \frac{12}{n}\sum_{k=1}^{K}\sum_{i=1}^{n_k}\frac{i}{n_k}Q_{(i)}^{(k)}.$$

*The computational cost of it is $O(n\log n)$, and it is consistent*.

## 2.5 Asymptotic Analysis

After giving the estimation of label projection correlation $\widehat{LPC}_n(X,Y)$, we present the asymptotic properties of it.

***Theorem 4.*** *Suppose $Y$ has a fixed categorical number $K$, the estimation of label projection correlation has the following asymptotic properties.*

*(1) Under the condition that $X$ and $Y$ are not independent.*

$$\begin{aligned}&\sqrt{n}\left(\widehat{LPCov}_n(X,Y) - LPCov(X,Y)\right)\xrightarrow{D} N\left(0,\sigma^2\right),\\&\sqrt{n}\left(\widehat{LPC}_n(X,Y) - LPC(X,Y)\right)\xrightarrow{D} N\left(0,\frac{\tilde{\sigma}^2}{\Delta^2}\right),\end{aligned} \tag{23}$$

*where $\Delta$ equals $S_1 + \pi S_3$, the expression and the consistent estimator of $\sigma^2$ and $\tilde{\sigma}^2$ are given in the Appendix. Besides, $\xrightarrow{D}$ is the symbol of converging in distribution.*

*(2) Under the condition that $X$ and $Y$ are independent.*

$$\begin{aligned}&n\widehat{LPCov}_n(X,Y)\xrightarrow{D}\sum_{i=1}^{\infty}\lambda_i\chi_i^2,\\&n\widehat{LPC}_n(X,Y)\xrightarrow{D}\frac{1}{\Delta}\sum_{i=1}^{\infty}\lambda_i\chi_i^2,\end{aligned} \tag{24}$$

*where $\{\chi_i^2\}_{i=1}^{\infty}$ is a series of random variable following $\chi^2(1)$ distribution independently, $\{\lambda_i\}_{i=1}^{\infty}$ is a series of constants such that $\sum_{i=1}^{\infty}\lambda_i = (K-1)\Delta$.*

The proofs of Theorem 4 are provided in the Appendix.

# 3. Numerical Experiments

In this section, we assess the effectiveness of our method through two simulated experiments and two real-world datasets. Before experiments, we first introduce the comparison methods. In addition to the Gini distance correlation (denoted by gCor) in Eq.(1), and the RKHS-based Gini distance correlation (denoted by gKCor) in Eq.(2), where the Mercer Kernel $\kappa(x,y)$ is taken as the Gaussian kernel $\exp(-\frac{\|x-y\|^2}{\sigma^2})/2$ with parameter $\sigma^2 = 1$, we also consider Pearson correlation shown as follows:

$$pCor(X,Y)=\frac{\sum_{i=1}^{p}\sum_{k=1}^{K}\left(p_k\mu_{ik}^2-\mu_i^2\right)}{\sum_{i=1}^{p}\sigma_i^2}, \tag{25}$$

where $p$ is the dimension of $X$, $\mu_i$ and $\sigma_i^2$ are the mean and variance of the *i-th* dimension of $X$ respectively, $\mu_{ik}$ is the mean of the *i-th* dimension of $X^{(k)}$.

### *3.1 Simulation data experiment*

We use the minimum model size, which includes all active variables, to assess the effectiveness of each feature selection method. The simulation experiments are repeated $M$ times, with the median and standard deviation of the minimum model size denoted as MMS and SD, respectively. Moreover, the proportion including a single active variable $X_i$, denoted by $P_i$, and the proportion including all active variables, denoted by $P_{all}$, are computed for a given model size $d=\lfloor n/\log n\rfloor$, where $n$ is the sample size and $\lfloor x\rfloor$ denotes the integer part of $x$.

***Example 1. Group Feature Selection***

In this example, we consider a group feature selection problem in Sang and Dang (2024). For each observation, the categorical response $Y$ is generated from three different distributions of three categories with the probability $P=(p_1,p_2,p_3)$ where $p_k=P(Y=k)$: (I) Balanced case $P_1=(\frac{1}{3},\frac{1}{3},\frac{1}{3})$; (II) Slightly unbalanced case $P_2=(\frac{3}{12},\frac{4}{12},\frac{5}{12})$; (III) Heavily unbalanced case $P_3=(\frac{1}{10},\frac{3}{10},\frac{3}{5})$. Given $Y=k$, $X=(X_{(1)},X_{(2)},\ldots,X_{(N)})$ is generated by $X=\mu^{(k)}+\varepsilon$, where $\mu^{(k)}=\left(\boldsymbol{\mu}_1^{(k)},\boldsymbol{\mu}_2^{(k)},\mathbf{0},\ldots,\mathbf{0}\right)$ and $\varepsilon$ is the error term. Let $\boldsymbol{\mu}^{(1)}=(1,0,0)^T$, $\boldsymbol{\mu}^{(2)}=(0,1,0)^T$, $\boldsymbol{\mu}^{(3)}=(0,0,1)^T$, and we consider two cases. In Case(a), $\boldsymbol{\mu}_1^{(k)}=\boldsymbol{\mu}_2^{(k)}=1.5\boldsymbol{\mu}^{(k)}$, which represents that $Y$ has the same role to $X_{(1)}$ and $X_{(2)}$. In Case(b), $\boldsymbol{\mu}_1^{(k)}=2\boldsymbol{\mu}^{(k)}$ and $\boldsymbol{\mu}_2^{(k)}=\boldsymbol{\mu}^{(k)}$, which represents that $Y$ has a greater effect on $X_{(1)}$. Besides, we consider three different distributions that error term $\varepsilon$ follows: (I) $N(0,1)$ distribution, which is one of the most common noises; (II) $t(1)$ distributions, which has a high probability of producing extreme values; (III) $t(2)$ distributions, which produces extreme values with a probability smaller than $t(1)$ but larger than $N(0,1)$. We set the sample size $n$ is 100, the group number $N$ is 400, and the experiments are repeated 60 times. The results presented in Table 1 demonstrate that our method consistently achieves the highest values for $P_1$, $P_2$, and $P_{all}$ across all experimental groups, while maintaining the lowest MMS and SD. These findings indicate the superior effectiveness of our approach compared to the other three methods.

**Table 1. Results of Example 1.**

| P | Noise | Index | Case(a) | | | | Case(b) | | | |
|---|---|---|---|---|---|---|---|---|---|---|
| | | | LPC | gCor | gKCor | Pearson | LPC | gCor | gKCor | Pearson |
| I | N(0,1) | $P_1$ | 1 | 1 | 1 | 1 | 1 | 1 | 1 | 1 |
| | | $P_2$ | 1 | 1 | 1 | 1 | 1 | 1 | 1 | 1 |
| | | $P_{all}$ | 1 | 1 | 1 | 1 | 1 | 1 | 1 | 1 |
| | | MMS | 2 | 2 | 2 | 2 | 2 | 2 | 2 | 2 |
| | | SD | 0 | 0 | 0 | 0 | 0 | 0 | 0 | 0 |
| | t(1) | $P_1$ | 0.983 | 0.417 | 0.817 | 0.050 | 1 | 0.700 | 0.933 | 0.100 |
| | | $P_2$ | 0.967 | 0.367 | 0.850 | 0.133 | 0.667 | 0.150 | 0.517 | 0.117 |
| | | $P_{all}$ | 0.950 | 0.167 | 0.683 | 0 | 0.667 | 0.117 | 0.483 | 0 |
| | | MMS | 2 | 74 | 8 | 273.500 | 11 | 110 | 23.500 | 285.500 |
| | | SD | 5.839 | 101.256 | 36.665 | 96.387 | 50.242 | 110.967 | 57.713 | 93.445 |
| | t(2) | $P_1$ | 1 | 1 | 0.983 | 0.900 | 1 | 1 | 1 | 0.950 |
| | | $P_2$ | 1 | 1 | 1 | 0.933 | 0.967 | 0.950 | 0.850 | 0.783 |
| | | $P_{all}$ | 1 | 1 | 0.983 | 0.850 | 0.967 | 0.950 | 0.850 | 0.750 |
| | | MMS | 2 | 2 | 2 | 2 | 2 | 2 | 3.500 | 4 |
| | | SD | 0 | 0 | 9.197 | 63.227 | 5.921 | 8.536 | 34.054 | 83.746 |
| II | N(0,1) | $P_1$ | 1 | 1 | 1 | 1 | 1 | 1 | 1 | 1 |
| | | $P_2$ | 1 | 1 | 1 | 1 | 1 | 1 | 1 | 1 |
| | | $P_{all}$ | 1 | 1 | 1 | 1 | 1 | 1 | 1 | 1 |
| | | MMS | 2 | 2 | 2 | 2 | 2 | 2 | 2 | 2 |
| | | SD | 0 | 0 | 0 | 0 | 0 | 0 | 0 | 0 |
| | t(1) | $P_1$ | 1 | 0.367 | 0.867 | 0.100 | 1 | 0.700 | 0.950 | 0.150 |
| | | $P_2$ | 0.983 | 0.467 | 0.800 | 0.100 | 0.700 | 0.150 | 0.50 | 0.083 |
| | | $P_{all}$ | 0.983 | 0.133 | 0.683 | 0 | 0.700 | 0.117 | 0.467 | 0 |
| | | MMS | 2 | 69.5 | 10.500 | 270 | 5.500 | 106 | 23 | 273.500 |
| | | SD | 8.310 | 103.774 | 29.858 | 97.634 | 48.808 | 111.202 | 65.347 | 92.043 |
| | t(2) | $P_1$ | 1 | 1 | 0.983 | 0.883 | 1 | 1 | 1 | 0.950 |
| | | $P_2$ | 1 | 1 | 1 | 0.917 | 0.967 | 0.933 | 0.817 | 0.700 |
| | | $P_{all}$ | 1 | 1 | 0.983 | 0.817 | 0.967 | 0.933 | 0.817 | 0.667 |
| | | MMS | 2 | 2 | 2 | 2 | 2 | 2 | 3 | 4.500 |
| | | SD | 0 | 0.406 | 4.755 | 55.016 | 5.034 | 12.107 | 29.380 | 57.489 |
| III | N(0,1) | $P_1$ | 1 | 1 | 1 | 1 | 1 | 1 | 1 | 1 |
| | | $P_2$ | 1 | 1 | 1 | 1 | 1 | 1 | 0.983 | 1 |
| | | $P_{all}$ | 1 | 1 | 1 | 1 | 1 | 1 | 0.983 | 1 |
| | | MMS | 2 | 2 | 2 | 2 | 2 | 2 | 2 | 2 |
| | | SD | 0 | 0.830 | 0.129 | 0.573 | 0 | 3.107 | 5.962 | 3.432 |
| | t(1) | $P_1$ | 0.933 | 0.200 | 0.650 | 0.083 | 1 | 0.317 | 0.867 | 0.083 |
| | | $P_2$ | 0.917 | 0.150 | 0.633 | 0.067 | 0.617 | 0.083 | 0.400 | 0.067 |
| | | $P_{all}$ | 0.867 | 0.017 | 0.400 | 0 | 0.617 | 0.017 | 0.350 | 0 |
| | | MMS | 4 | 135.500 | 36.500 | 251.500 | 10 | 137 | 45.500 | 250.500 |
| | | SD | 21.350 | 108.126 | 51.809 | 102.068 | 78.357 | 107.441 | 97.456 | 103.128 |
| | t(2) | $P_1$ | 1 | 0.967 | 0.983 | 0.633 | 1 | 1 | 1 | 0.833 |
| | | $P_2$ | 1 | 0.950 | 1 | 0.633 | 0.933 | 0.600 | 0.767 | 0.317 |
| | | $P_{all}$ | 1 | 0.933 | 0.983 | 0.400 | 0.933 | 0.600 | 0.767 | 0.267 |
| | | MMS | 2 | 8 | 2 | 27 | 2 | 17 | 4 | 33 |
| | | SD | 0.129 | 6.525 | 14.339 | 48.562 | 7.467 | 19.340 | 31.233 | 73.531 |

***Example 2. Genome-Wide Association Studies***

In this example, we consider a Genome-Wide Association Study (GWAS) example mentioned by Cui et al. (2015). In GWAS, researchers usually collect genetic data containing an extremely large number of single-nucleotide polymorphisms (SNPs). Generally, the SNPs are usually classified into three categories denoted by $AA$, $Aa$ and $aa$. We denote $Z_{ij}$ as the indicators of the dominant effect of the *j-th* SNP for *i-th* subject and generate it in the following way

$$Z_{ij} = \begin{cases} 1, & \text{if} \quad X_{ij} < q_{1j} \\ 0, & \text{if} \quad q_{1j} \le X_{ij} < q_{3j}, \\ -1, & \text{if} \quad X_{ij} \ge q_{3j} \end{cases}$$

where $X_i = \left(X_{i1}, X_{i2}, ..., X_{ip}\right) \sim N(0,\Sigma)$ for $i = 1,\dots,n$, $\Sigma = (\rho_{ij})_{p\times p}$ with $\rho_{ij} = 0.5^{|i-j|}$, and $q_{1j}$, $q_{3j}$ are the first and third quartiles of each $\{X_{ij}\}_{i=1}^{n}$, respectively. The response variable $Y$ is generated by

$$Y = \beta_1 Z_1 + \beta_2 Z_2 + 2\beta_3 Z_{10} + 2\beta_4 Z_{20} - 2\beta_5 |Z_{100}| + \varepsilon,$$

where $\beta_i = (-1)^U (2\log(n)/\sqrt{n} + |Z|)$ for $i = 1,\dots,5$ with $U \sim Bernoulli(0.4)$ and $Z$ are i.i.d from $N(0,1)$. The error term $\varepsilon$ follows $N(0,1)$, $t(1)$ and $t(2)$ distribution, respectively. In this model, $Z_1$, $Z_2$, $Z_{10}$, $Z_{20}$, $Z_{100}$ are five active SNPs. We set the sample size $n = 200$, the variable number $p = 2000$, and the procedure is repeated 100 times. The results are shown in Table 2. To illustrate better, the maximum value of $P_i$ and $P_{all}$ for each column under the certain error term is bolded and the second largest value is underlined. The first and second smallest values of MMS and SD are processed in the same manner.

From Table 2, it can be concluded that when the error term follows $N(0,1)$, the results of Pearson correlation are relatively better. When the error term follows $t(1)$, our method performs better than the other three methods. When the error term follows $t(2)$, our method and Gini distance correlation exhibit the optimal performance in $P_i$ and $P_{all}$. In summary, when the error term is more likely to produce extreme values, employing our method for feature selection leads to better results.

**Table 2. Results of Example 2.**

| ε | Method | $P_1$ | $P_2$ | $P_{10}$ | $P_{20}$ | $P_{100}$ | $P_{all}$ | MMS | SD |
|---|---|---|---|---|---|---|---|---|---|
| N(0,1) | LPC | 0.68 | 0.62 | 0.99 | **0.99** | 0.88 | 0.41 | 132.0 | 621.4 |
| | gCor | 0.68 | 0.63 | 1.00 | **0.99** | 0.92 | 0.45 | 104.5 | 605.1 |
| | gKCor | 0.31 | 0.34 | 0.83 | 0.81 | 0.58 | 0.07 | 682.0 | **537.8** |
| | Pearson | **0.71** | **0.65** | **1.00** | **0.99** | **0.93** | **0.47** | **51.5** | 582.1 |
| t(1) | LPC | **0.52** | **0.52** | **0.96** | **0.92** | **0.74** | **0.19** | **331.5** | 569.0 |
| | gCor | 0.46 | 0.48 | 0.92 | 0.90 | 0.68 | 0.16 | 334.0 | **568.8** |
| | gKCor | 0.30 | 0.34 | 0.76 | 0.69 | 0.54 | 0.03 | 610.0 | 633.3 |
| | Pearson | 0.18 | 0.22 | 0.35 | 0.48 | 0.26 | 0.01 | 1419.5 | 601.9 |
| t(2) | LPC | **0.56** | **0.59** | **0.98** | **0.98** | **0.92** | 0.28 | 318.5 | 604.6 |
| | gCor | **0.56** | 0.56 | **0.98** | **0.98** | **0.92** | **0.29** | 304.0 | 606.2 |
| | gKCor | 0.31 | 0.25 | 0.80 | 0.78 | 0.52 | 0.03 | 799.5 | **572.1** |
| | Pearson | 0.55 | 0.53 | **0.98** | 0.96 | 0.90 | 0.28 | **243.5** | 619.1 |

### *3.3 Real data analysis*

We compare our method in feature selection and classification using two real-world datasets: the DrivFace dataset for image data and the Arcene dataset for cancer mass-spectrometric data. Both datasets are available from the UCI Machine Learning Repository.

***Example 3. Image data analysis***

The DrivFace dataset includes image sequences of individuals driving in real-world scenarios. It comprises 606 samples collected from four drivers, consisting of two females and two males. The drivers' faces are standardized to 80x80 pixels, and their associated gaze directions are labeled as "looking right",

"frontal", and "looking left". We aim to identify the pixels strongly related to the gaze direction. First, 70% of the samples are randomly selected to constitute the training set, while the remaining samples are allocated to the test set. Subsequently, the correlation coefficients between each pixel and the gaze direction are computed within the training dataset, as depicted in Fig.1. For a more detailed comparison, Fig.1 also presents the 500 pixels with the largest correlation coefficients under different methods. The results indicate that for our method, the pixels with higher correlation coefficients are more concentrated on the facial region, which illustrates that our method captures the facial features of drivers with different gaze directions better than other methods.

To demonstrate the effectiveness of our method, we use highly correlated pixels to classify them on the test set and compare the result to the true gaze direction. We utilize the random forest with 100 decision trees as the classifier and compute the average classification accuracy over 10 independent experiments to avoid the impact of randomness. Table 3 presents the test accuracy using the top $d$ pixels with the highest correlation coefficients for each method. The results indicate that our method performs best in all cases, validating its effectiveness in feature selection of image data.

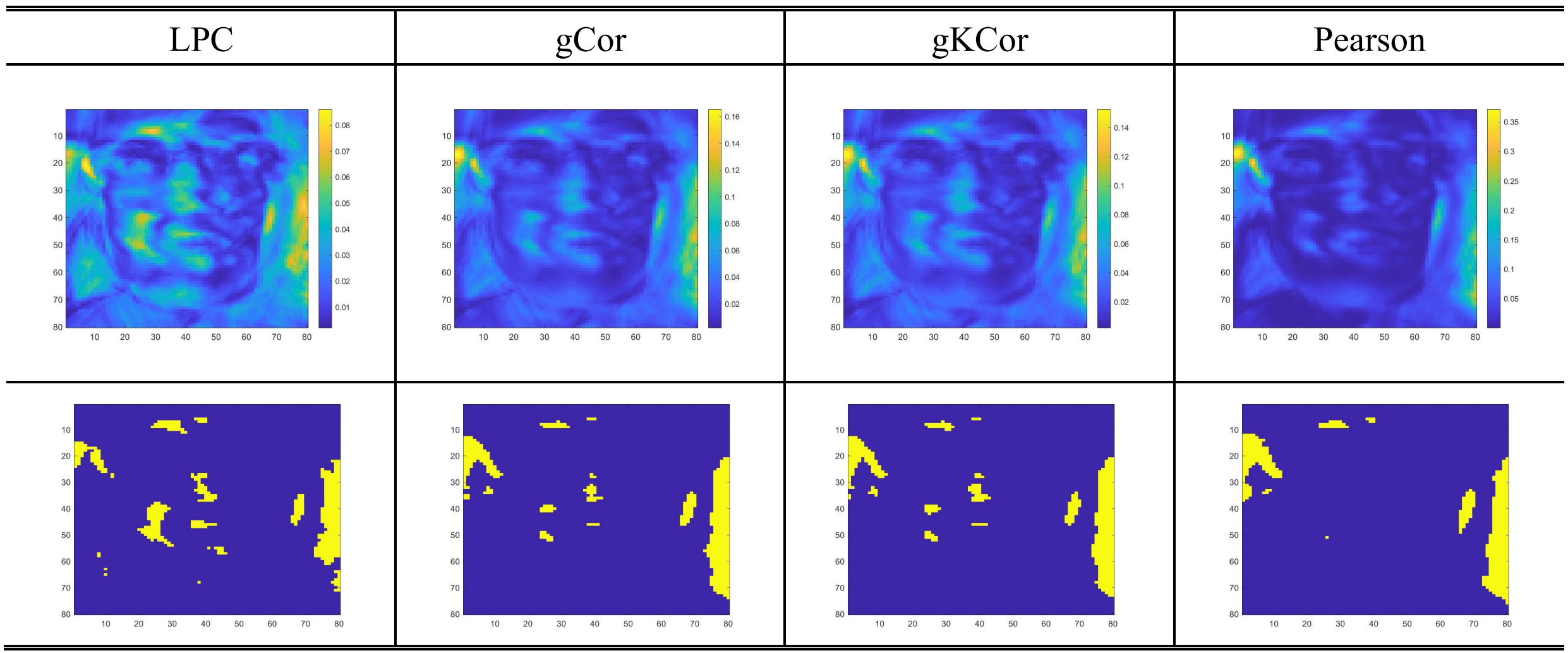

**Fig.1 Correlation coefficients of each pixel and the most correlated 500 pixels with different methods.**

**Table 3. Test accuracy based on the top $d$ pixels selected by each correlation.**

| Method | 100 | 200 | 300 | 400 | 500 | 600 |
|---|---|---|---|---|---|---|
| LPC | **0.9626** | **0.9632** | **0.9665** | **0.9670** | **0.9698** | **0.9698** |
| gCor | 0.9604 | 0.9560 | 0.9566 | 0.9615 | 0.9681 | 0.9687 |
| gKCor | 0.9577 | 0.9560 | 0.9577 | 0.9626 | 0.9670 | 0.9687 |
| Pearson | 0.9571 | 0.9560 | 0.9582 | 0.9560 | 0.9560 | 0.9571 |

***Example 4. Cancer feature selection and classification***

The ARCENE dataset is constructed by merging three mass spectrometry datasets and is designed to distinguish cancer versus normal patterns from mass-spectrometric data. We use the training set and validation set from the dataset, each containing 100 samples, of which 44 are cancer samples and 56 are normal samples. Each sample includes 10,000 features, with 7,000 predictive features and 3,000 non-predictive "probe" features. The order of features and patterns in the dataset has been randomized. To build the classification model, we first calculate the correlation between each feature and the patterns in the training set, select $d$ features with the highest correlation coefficient, and use these features to classify the samples in the validation set. The random forest with 100 decision trees is still applied as the classifier, and each classification experiment is repeated 10 times independently. The test accuracy results are presented in Table 4.

**Table 4. Test accuracy of Arcene data based on the top $d$ features selected by each correlation.**

| Method | 20 | 40 | 60 | 80 | 100 | 200 | 300 | 400 | 500 | 600 |
|---|---|---|---|---|---|---|---|---|---|---|
| LPC | **0.722** | **0.744** | **0.742** | **0.735** | **0.731** | 0.714 | 0.723 | 0.729 | 0.744 | 0.738 |
| gCor | 0.590 | 0.622 | 0.701 | 0.714 | 0.707 | 0.710 | 0.722 | 0.724 | 0.716 | 0.713 |
| gKCor | 0.690 | 0.725 | 0.724 | 0.713 | 0.715 | 0.705 | **0.733** | **0.757** | **0.772** | **0.760** |
| Pearson | 0.619 | 0.609 | 0.612 | 0.679 | 0.693 | **0.717** | 0.722 | 0.736 | 0.753 | 0.747 |

From Table 4, the label projection correlation and RKHS-based Gini distance correlation, which do not have limitations on the variables, can achieve higher classification accuracy. Additionally, the classification accuracies of label projection correlation are consistently greater than those of Gini distance correlation. These two results may be attributed to the fact that most of the features in the dataset exhibit heavy-tailed distributions. Specifically, in the 10000 variables, 6845 have a kurtosis value exceeding 3, and 5185 have a kurtosis value exceeding 5. This may cause the other two methods to fail in selecting predictive features due to their moment limitations not being satisfied. Moreover, when the number of selected features is relatively small, our method shows a significantly higher classification accuracy compared to the other methods. This indicates that the highly correlated features selected by our method are more effective in distinguishing different patterns.

## 4. Conclusion

In this paper, the label projection correlation is introduced for measuring the dependence between numerical and categorical variables. We provide its definition, statistical properties, estimation, and demonstrate some excellent properties of it when the numerical variable is one-dimensional. The asymptotic theorems of the estimation are also derived. Subsequently, our method is compared with the other three metrics on two simulated datasets, which shows that our method is effective in group variable selection and performs better when the error term follows the distribution with a higher probability of producing extreme values. In image data analysis, the proposed correlation can better capture the facial features of drivers with different gaze directions, and achieve the highest classification accuracy with selected pixels. In cancer mass-spectrometric data analysis, where most features exhibit heavy-tailed distributions, the highly correlated features selected by our approach demonstrate better classification ability.

However, the novel metric still has some limitations. For example, in feature selection, inappropriate selection may occur due to the correlation between features. Therefore, measures of conditional independence should be considered in future work to address this issue. Moreover, the computational cost of our correlation estimation in multi-dimensional case is $O(n^3)$, which is much higher than that of the other metrics. Although we can reduce it by random sampling or some other methods, the convergence and effectiveness of these algorithms are new challenges.

## Acknowledgment

The financial support from the funds of the National Natural Science Foundation of China (62171018) is gratefully acknowledged.

## Appendix

The Appendix contains the proof of Theorem 1-4 and Lemma 3.

***The proof of Theorem 1.*** (1) and (2) in Theorem 1 are obvious. Next, we will prove (3). According to Lemma 2, Eq.(8) and Eq.(9), when $LPC(X,Y)=1$, for all $\boldsymbol{\alpha} \in \mathbb{R}^p$, we have

$$E_{X_1}\left[E_Y\left[Var_X\left[I\left(\boldsymbol{\alpha}^T X \le \boldsymbol{\alpha}^T X_1\right) | Y\right]\right]\right] = 0.$$

As variance is greater than or equal to zero, for any $Y$ and $X_1$, there will be

$$\begin{aligned} 0 &= Var_X\left[I\left(\boldsymbol{\alpha}^T X \le \boldsymbol{\alpha}^T X_1\right) | Y\right] \\ &= E_X\left[I\left(\boldsymbol{\alpha}^T X \le \boldsymbol{\alpha}^T X_1\right) | Y\right]\left(1 - E_X\left[I\left(\boldsymbol{\alpha}^T X \le \boldsymbol{\alpha}^T X_1\right) | Y\right]\right). \end{aligned}$$

Therefore, $E_X[I(\boldsymbol{\alpha}^T X \le \boldsymbol{\alpha}^T X_1) | Y]$ can only be equal to 0 or 1, which is equivalent to the condition that for all $k$, $X^{(k)}$ is equal to a constant vector $\boldsymbol{c_k}$. Besides, when all $\boldsymbol{c_k}$ are the same, $X$ and $Y$ are independent. Removing this case, (3) holds.

According to the definition of $PCor(X,Y)$, (4) is obvious. Next, we prove (5). When $X$ is a continuous random variable, we have

$$\begin{aligned} LPCov_{upper}(X,Y) &= \frac{1}{c_p}\int_{\|\boldsymbol{\alpha}\|=1} E_{X_1}\left[Var_X\left(I\left(\boldsymbol{\alpha}^T X \le \boldsymbol{\alpha}^T X_1\right)\right)\right] d\boldsymbol{\alpha} \\ &= \frac{1}{c_p}\int_{\|\boldsymbol{\alpha}\|=1} E_{X_1}\left[E_X\left(I\left(\boldsymbol{\alpha}^T X \le \boldsymbol{\alpha}^T X_1\right)\right) - \left(E_X\left(I\left(\boldsymbol{\alpha}^T X \le \boldsymbol{\alpha}^T X_1\right)\right)\right)^2\right] d\boldsymbol{\alpha} \\ &= \frac{1}{c_p}\int_{\|\boldsymbol{\alpha}\|=1} E_{U_1}\left[E_U\left(I\left(U \le U_1\right)\right) - \left(E_U\left(I\left(U \le U_1\right)\right)\right)^2\right] d\boldsymbol{\alpha} \\ &= \frac{1}{c_p}\int_{\|\boldsymbol{\alpha}\|=1}\int_0^1 \left(x - x^2\right) dx d\boldsymbol{\alpha} = \frac{\pi}{3}, \end{aligned}$$

where $U_1$ is the independent copy of $U$. Since $S_3$ equals zero when $X$ is continuous, $S_1$ equals $\pi/3$. Therefore, $LPC(X,Y) = (\pi - 3S_2)/\pi$.

Finally, we prove (6). By the definition of $LPCov(X,Y)$,

$$\begin{aligned} LPCov(X,Y) = &\frac{1}{c_p}\Bigg[\sum_{i=1}^{K} p_i E[\int_{\|\boldsymbol{\alpha}\|=1} I(\boldsymbol{\alpha}^T(X_1 - X_3) \le 0) I(\boldsymbol{\alpha}^T(X_2 - X_3) \le 0) \mathrm{d}\boldsymbol{\alpha} \mid Y_1 = Y_2 = i] \\ &- E[\int_{\|\boldsymbol{\alpha}\|=1} I(\boldsymbol{\alpha}^T(X_1 - X_3) \le 0) I(\boldsymbol{\alpha}^T(X_2 - X_3) \le 0) \mathrm{d}\boldsymbol{\alpha}]\Bigg] \\ = &\frac{1}{c_p}\sum_{i=1}^{K}\sum_{j=1}^{K} p_i p_j E[\int_{\|\boldsymbol{\alpha}\|=1} I(\boldsymbol{\alpha}^T(X_1 - X_3) \le 0) I(\boldsymbol{\alpha}^T(X_2 - X_3) \le 0) \mathrm{d}\boldsymbol{\alpha} \mid Y_1 = Y_2 = i, Y_3 = j] - \frac{2}{3}\pi, \end{aligned} \tag{26}$$

where the second part of Eq.(26) can be derived through Lemma 1 and $S_1 = \pi/3$. Next, we will derive the first part of Eq.(26), and an important lemma is presented here for the following derivation.

***Lemma 4**. (Gupta 1963) Let $(Z_1, Z_2)$ be the bivariate normal distribution with mean zero and*

*correlation* $\rho$, *Then,* $E[I(Z_1 \le 0)I(Z_2 \le 0)] = 1/4 + (2\pi)^{-1}\arcsin\rho$.

For arbitrary unit vector $\boldsymbol{\alpha} \in \mathbb{R}^p$, when $Y_1 = Y_2 = i$ and $Y_3 = k$, $\boldsymbol{\alpha}^T(X_1 - X_3)$ and $\boldsymbol{\alpha}^T(X_2 - X_3)$ jointly follow the bivariate normal distribution such that

$$\begin{pmatrix} \boldsymbol{\alpha}^T(X_1 - X_3) \\ \boldsymbol{\alpha}^T(X_2 - X_3) \end{pmatrix} \sim N\left( \begin{pmatrix} 0 \\ 0 \end{pmatrix}, \begin{pmatrix} \sigma_i^2 + \sigma_j^2 & \sigma_j^2 \\ \sigma_j^2 & \sigma_i^2 + \sigma_j^2 \end{pmatrix} \right).$$

Hence, correlation $\rho = \sigma_j^2 / (\sigma_i^2 + \sigma_j^2)$. By Lemma 4, we have:

$$\frac{1}{c_p}\sum_{i=1}^{K}\sum_{j=1}^{K} p_i p_j E[\int_{\|\boldsymbol{\alpha}\|=1} I(\boldsymbol{\alpha}^T(X_1 - X_3) \le 0) I(\boldsymbol{\alpha}^T(X_2 - X_3) \le 0)\mathrm{d}\boldsymbol{\alpha} \mid Y_1 = Y_2 = i, Y_3 = j]$$

$$= 2\pi \sum_{i=1}^{K}\sum_{j=1}^{K} p_i p_j \left( \frac{1}{4} + \frac{1}{2\pi}\arcsin\left( \frac{\sigma_j^2}{\sigma_i^2 + \sigma_j^2} \right) \right) = \frac{\pi}{2} + \frac{1}{2}\sum_{i=1}^{K}\sum_{j=1}^{K} p_i p_j \left( \arcsin\left( \frac{\sigma_i^2}{\sigma_i^2 + \sigma_j^2} \right) + \arcsin\left( \frac{\sigma_j^2}{\sigma_i^2 + \sigma_j^2} \right) \right).$$

Combining the result of $LPCov(X,Y)$ and $LPCov_{upper}(X,Y) = \pi/3$ when $X$ is continuous, the expression of $LPC(X,Y)$ can be derived. The proof of Theorem 1 is complete.

□

***The proof of Theorem 2.*** First, $\hat{S}_2$ can be expressed as follows:

$$\hat{S}_2 = \sum_{k=1}^{K} \frac{1}{\hat{p}_k} \frac{1}{n^3} \sum_{i,j,l} a_{ijl} I(Y_i = Y_j = k) = \sum_{k=1}^{K} \frac{1}{\hat{p}_k} \frac{1}{n^3} \sum_{i,j,l} \frac{1}{3!} \sum\nolimits_3 \{a_{ijl} I(Y_i = Y_j = k)\},$$

where $\Sigma_3$ is the permutation summation of three distinct elements $(i, j, l)$. By the strong convergence theorem of V-statistic(Serfling 1980) and the continuous mapping theorem, $\hat{S}_2$ converges to $S_2$ almost surely. Besides, $\hat{S}_1$ and $\hat{S}_3$ converge to $S_1$ and $S_3$, respectively. By the continuous mapping theorem, $\widehat{LPC_n}(X,Y)$ converges to $LPC(X,Y)$ almost surely. The proof is completed.

□

***The proof of Lemma 3.*** For $E(|Q_1 - Q_2|)$, we have

$$\begin{aligned} E(|Q_1 - Q_2|) &= E\left( \left| I(X_3 \le X_1) - I(X_3 \le X_2) \right| \right) \\ &= E\left( \left| I(X_3 \le X_1) - I(X_3 \le X_2) \right|^2 \right) \\ &= E\left( I(X_3 \le X_1) + I(X_3 \le X_2) - 2I(X_3 \le X_1) I(X_3 \le X_2) \right) \\ &= 2E_{X_2}\left( Var_{X_1}\left( I(X_1 \ge X_2) \right) \right). \end{aligned}$$

For $\sum_{k=1} p_k E\left( \left| Q_1^{(k)} - Q_2^{(k)} \right| \right)$, we have

$$\begin{aligned} \sum_{k=1} p_k E\left( \left| Q_1^{(k)} - Q_2^{(k)} \right| \right) &= \sum_{k=1} p_k E\left( \left| I\left(X_3 \le X_1^{(k)}\right) - I\left(X_3 \le X_2^{(k)}\right) \right| \right) \\ &= \sum_{k=1} p_k E\left( \left| I\left(X_3 \le X_1^{(k)}\right) - I\left(X_3 \le X_2^{(k)}\right) \right|^2 \right) \\ &= \sum_{k=1} p_k E\left( I\left(X_3 \le X_1^{(k)}\right) + I\left(X_3 \le X_2^{(k)}\right) - 2I\left(X_3 \le X_1^{(k)}\right) I\left(X_3 \le X_2^{(k)}\right) \right) \\ &= 2E_{X_2}\left( E_{Y_1}\left( Var_{X_1}\left( I(X_1 \ge X_2) \mid Y_1 \right) \right) \right). \end{aligned}$$

The proof of Lemma 3 is complete.

$\square$

***The proof of Theorem 3.***

(1) As $X$ is continuous, $X^{(k)}$ is also continuous. Hence, we have

$$Var_Y\left[E_X\left[I(X\le t)\,|\,Y\right]\right]=Var_Y\left[1-E_X\left[I(X\le t)\,|\,Y\right]\right]=Var_Y\left[E_X\left[I(X\ge t)\,|\,Y\right]\right],$$

$$\begin{aligned}Var_X\left(I(X\le t)\right)&=E\left[I(X\le t)\right]\left(1-E\left[I(X\le t)\right]\right)\\&=\left(1-E\left[I(X\ge t)\right]\right)E\left[I(X\ge t)\right]=Var_X\left(I(X\ge t)\right).\end{aligned}$$

Therefore, $T_1 = {}^{-}T_1$ and $T_2 = {}^{-}T_2$ hold, and $LPC(X,Y)$ can be expressed as follows:

$$LPC(X,Y)=\frac{T_1-T_2}{T_1}.$$

Then we explore methods for simplifying $T_1$ and $T_2$. First, $T_1$ can be computed as follows:

$$\begin{aligned}T_1&=2\int_0^1\int_0^{x_2}(x_2-x_1)p(x_1)p(x_2)dx_1dx_2\\&=2\int_0^1\int_0^{x_2}(x_2-x_1)dx_1dx_2=\frac{1}{3},\end{aligned}$$

where $p(x)$ refers to the probability density function (p.d.f) of the uniform distribution over $[0,1]$. Next, we provide the simple expression of $T_2$ as follows:

$$\begin{aligned}T_2&=2\sum_{k=1}^{K}p_k\int_0^1\int_0^{x_2}(x_2-x_1)q^{(k)}(x_1)q^{(k)}(x_2)dx_1dx_2\\&=2\sum_{k=1}^{K}p_k\left[2\int_0^1 xq^{(k)}(x)F_{Q^{(k)}}(x)dx-\int_0^1 xq^{(k)}(x)dx\right]\\&=2\sum_{k=1}^{K}p_k\left[2E(Q^{(k)}F_{Q^{(k)}}(Q^{(k)}))-E(Q^{(k)})\right]\\&=4\sum_{k=1}^{K}p_kE(Q^{(k)}F_{Q^{(k)}}(Q^{(k)}))-1,\end{aligned}$$

where $q^{(k)}(x)$ is denoted as the p.d.f of $Q^{(k)}$. Combining the result of $T_1$ and $T_2$, we have

$$LPC(X,Y)=\frac{\frac{4}{3}(1-3\sum_{k=1}^{K}p_kE(Q^{(k)}F_{Q^{(k)}}(Q^{(k)})))}{\frac{1}{3}}=4(1-3\sum_{k=1}^{K}p_kE(Q^{(k)}F_{Q^{(k)}}(Q^{(k)}))),$$

where $F_{Q^{(k)}}(Q^{(k)})$ is equivalent to $F_{X^{(k)}}(X^{(k)})$.

(2) A decomposition of $\widehat{LPC}_n(X,Y)$ is demonstrated here to prove the consistency of $\widetilde{LPC}_n(X,Y)$. First, it is obvious that $\widehat{T}_1 = {}^{-}\hat{T}_1$ and $\widehat{T}_2 = {}^{-}\hat{T}_2$ when $X$ is continuous, and

$$\hat{T}_1=\frac{2}{n^2}\sum_{i=1}^{n}(2i-n-1)Q_{(i)}=\frac{2}{n^2}\sum_{i=1}^{n}(2i-n-1)\frac{i}{n}=\frac{2}{n^3}\left[\frac{n(n+1)(2n+1)}{3}-(n+1)\frac{n(n+1)}{2}\right]=\frac{1}{3}-\frac{1}{3n^2},$$

$$
\begin{aligned}
\hat{T}_2 &= \frac{2}{n}\sum_{k=1}^{K}\frac{1}{n_k}\sum_{i=1}^{n_k}\left(2i-n_k-1\right)Q_{(i)}^{(k)} = \frac{2}{n}\sum_{k=1}^{K}\frac{1}{n_k}\sum_{i=1}^{n_k}\left(2i-1\right)Q_{(i)}^{(k)} - \frac{2}{n}\sum_{i=1}^{n}\frac{i}{n} \\
&= \frac{4}{n}\sum_{k=1}^{K}\sum_{i=1}^{n_k}\frac{i}{n_k}Q_{(i)}^{(k)} - \frac{2}{n}\sum_{k=1}^{K}\frac{1}{n_k}\sum_{i=1}^{n_k}Q_{(i)}^{(k)} - \frac{n+1}{n} = \frac{4}{n}\sum_{k=1}^{K}\sum_{i=1}^{n_k}\frac{i}{n_k}Q_{(i)}^{(k)} - 1 + O(n^{-1}).
\end{aligned}
$$

Combing the above expression of $\hat{T}_1$ and $\hat{T}_2$, $\widehat{LPC}_n(X,Y)$ can be decomposed as:

$$
\widehat{LPC}_n(X,Y) = \frac{\widehat{T}_1-\widehat{T}_2}{\widehat{T}_1} = \frac{\widehat{T}_1-\widehat{T}_2}{1/3} + \left(\frac{\widehat{T}_1-\widehat{T}_2}{\widehat{T}_1} - \frac{\widehat{T}_1-\widehat{T}_2}{1/3}\right) = 3\left(\widehat{T}_1-\widehat{T}_2\right) + \frac{\widehat{T}_1-\widehat{T}_2}{\widehat{T}_1}\frac{1}{n^2} = \widetilde{LPC}_n(X,Y) + O(n^{-1}).
$$

Since $\widehat{LPC}_n(X,Y)$ converges to $LPC(X,Y)$ almost surely as $n \to \infty$, $\widetilde{LPC}_n(X,Y)$ is also consistent. The proof of Theorem 3 is complete.

□

***The proof of Theorem 4***.

(1) **When $X$ and $Y$ are dependent.** First, according to the proof of Theorem 3.3 in Cui and Zhong (2019), $\widehat{LPCov}_n(X,Y)$ can be simplified as follows:

$$
\begin{aligned}
&\widehat{LPCov}_n(X,Y) - LPCov(X,Y) = \sum_{i=1}^{n}\int_{\|\boldsymbol{\alpha}\|=1} I(\boldsymbol{\alpha}^T X_i, Y_i)\, d\boldsymbol{\alpha} + o_p(n^{-\frac{1}{2}}) \\
&= \sum_{i=1}^{n}\left[f_1(Y_i) - S_2\right] + 2\left[f_2(X_i) + f_3(X_i,Y_i) - (S_1 - S_2)\right] + \left[\left(f_4(X_i) - f_5(X_i)\right) - \left(S_1 - S_2\right)\right] + o_p(n^{-\frac{1}{2}}),
\end{aligned} \tag{27}
$$

where

$$
\begin{aligned}
I(X_i,Y_i) = \sum_{k=1}^{K}\{ &-\int F_{X|Y=k}^2(x)\,dF_X(x)(I\{Y_i = k\} - p_k) + 2\int\left(F_{X|Y=k}(x) - F_X(x)\right)\left(I\{X_i \le x, Y_i = k\} - F_{X|Y=k}(x)p_k\right)dF_X(x) \\
&+ p_k[[F_{X|Y=k}(X_i) - F_X(X_i)]^2 - \int\left(F_{X|Y=k}(x) - F_X(x)\right)^2 dF_X(x)]\},
\end{aligned}
$$

and

$$
\begin{aligned}
&f_1(Y) = E[a_{123} \mid Y_1 = Y_2 = Y], f_2(X) = E[a_{123} \mid X_1 = X], f_3(X,Y) = E[a_{123} \mid X_1 = X, Y_2 = Y], \\
&\qquad f_4(X) = E[a_{123} \mid X_3 = X], f_5(X) = \sum\nolimits_{k=1}^{K} p_k E[a_{123} \mid X_3 = X, Y_1 = Y_2 = k].
\end{aligned}
$$

By the central limit theorem, the asymptotic normality of $\widehat{LPCov}_n(X,Y)$ is obtained, and the asymptotic variance $\sigma^2$ is expressed as $\sigma^2 = Var[f_1(Y) + 2f_2(X) - 2f_3(X,Y) + f_4(X) - f_5(X)]$.

Next, we will show the asymptotic normality of $\widehat{LPC}_n(X,Y)$. First, we have the decomposition of $\widehat{LPC}_n(X,Y)$ shown below:

$$
\begin{aligned}
&\widehat{LPC}_n(X,Y) - LPC\left(X,Y\right) \\
&= \frac{\left(S_1 + \pi S_3\right)\left(\widehat{LPCov}_n(X,Y) - LPCov(X,Y)\right) - LPCov(X,Y)\left((\hat{S}_1 - S_1) + \pi(\hat{S}_3 - S_3)\right)}{\left(S_1 + \pi S_3\right)\left(\hat{S}_1 + \pi\hat{S}_3\right)}.
\end{aligned} \tag{28}
$$

Using the projection method shown in Serfling (1980), $\hat{S}_1 - S_1$, $\hat{S}_3 - S_3$ can be simplified as follows:

$$
\begin{aligned}
\hat{S}_1 - S_1 &= \frac{3}{n}\sum_{i=1}^{n}\left(E\left[\frac{1}{3!}\sum\nolimits_3 a_{ijl} \mid X_i, Y_i\right] - S_1\right) + o_p(n^{-\frac{1}{2}}) \\
&= \frac{1}{n}\sum_{i=1}^{n}\left(2f_2\left(X_i\right) + f_4\left(X_i\right) - 3S_1\right) + o_p(n^{-\frac{1}{2}}),
\end{aligned} \tag{29}
$$

$$\hat{S}_3 - S_3 = \frac{2}{n}\sum_{i=1}^{n}\left(f_6\left(X_i\right) - S_3\right) + o_p\left(n^{-\frac{1}{2}}\right), \tag{30}$$

where $\Sigma_3$ is the permutation summation of three distinct elements $(i,j,l)$ and $f_6\left(X_i\right) = E[I(X = X_i) \mid X_i]$. Hence, by the Slutsky's Theorem,

$$\sqrt{n}\left(\widehat{LPC}_n(X,Y)\left(X,Y\right) - LPC\left(X,Y\right)\right) \xrightarrow{D} N(0, \frac{\tilde{\sigma}^2}{\Delta^2}),$$

where $\tilde{\sigma}^2 = Var[f_1(Y) + 2(1 - LPC(X,Y))f_2(X) + 2f_3(X,Y) + (1 - LPC(X,Y))f_4(X) - f_5(X) - 2\pi LPC(X,Y)f_6(X)]$ , and $LPC(X,Y)$ is the correlation measure, not a random variable.

(2) **When $X$ and $Y$ are independent.** First, according to the Lemma A.1 in Cui and Zhong (2019), when $X$ and $Y$ are independent, $\widehat{LPCov}_n(X,Y)$ can be expressed as follows:

$$\begin{aligned}
\widehat{LPCov}_n(X,Y) &= \frac{1}{c_p}\sum_{k=1}^{K}\frac{1}{\hat{p}_k}\int_{\|\boldsymbol{\alpha}\|=1}\int\left(\frac{1}{n}\sum_{i=1}^{n}f_{\boldsymbol{\alpha},i}\left(x,k\right)\right)^2 d\hat{F}_{\boldsymbol{\alpha}}(x)d\boldsymbol{\alpha} \\
&= \frac{1}{c_p}\sum_{k=1}^{K}\frac{1}{p_k}\int_{\|\boldsymbol{\alpha}\|=1}\int\left(\frac{1}{n}\sum_{i=1}^{n}f_{\boldsymbol{\alpha},i}\left(x,k\right)\right)^2 d\hat{F}_{\boldsymbol{\alpha}}(x)d\boldsymbol{\alpha} + O_p\left(n^{-\frac{3}{2}}\right) \\
&= \sum_{k=1}^{K}\frac{1}{p_k}\frac{1}{n^3}\sum_{i,j,l=1}^{n}\left(-a_{ijl} + a_{i\bullet l} + a_{\bullet jl} - a_{\bullet\bullet l}\right)\left(I\left(Y_i = k\right) - p_k\right)\left(I\left(Y_j = k\right) - p_k\right) + O_p\left(n^{-\frac{3}{2}}\right) \\
&= \frac{1}{n^3}\sum_{i,j,l=1}^{n} b_{ijl} + O_p\left(n^{-\frac{3}{2}}\right),
\end{aligned} \tag{31}$$

where $\hat{F}_{\boldsymbol{\alpha}}(x) = \sum_{i=1}^{n} I(\boldsymbol{\alpha}^T X_i \leq x)\big/n$, $f_{\boldsymbol{\alpha},i}\left(x,k\right) = \left(I(\boldsymbol{\alpha}^T X_i \leq x) - F_{\boldsymbol{\alpha}^T X}(x)\right)\left(I(Y_i = k) - p_k\right)$, $a_{i\bullet l} = E[a_{ijl} \mid X_i, X_l]$, $a_{\bullet jl} = E[a_{ijl} \mid X_j, X_l]$, and $a_{\bullet\bullet l} = E[a_{ijl} \mid X_l]$. $b_{ijl}$ is defined in the obvious manner, and $E[b_{ijl}] = LPCov(X,Y)$ when $(i,j,l)$ are different. Then, we define a V-statistic $\hat{V}_n$ as follows:

$$\hat{V}_n = \frac{1}{n^3}\sum_{i,j,l=1}^{n}\frac{1}{3!}\sum\nolimits_3 b_{ijl} = \frac{1}{n^3}\sum_{i,j,l=1}^{n} b_{ijl}, \tag{32}$$

where $\Sigma_3$ is the permutation summation of three distinct elements $(i,j,l)$. For all $\{X_i, Y_i\}_{i=1}^{n}$, it is easy to verify that

$$E\left[\frac{1}{3!}\sum\nolimits_3 b_{ijl} \mid X_i, Y_i\right] = 0. \tag{33}$$

Hence, $\widehat{LPCov}_n(X,Y)$ does not show the asymptotic normality. However, we have:

$$\begin{aligned}
E\left[\frac{1}{3!}\sum\nolimits_3 b_{ijl} \mid X_i, Y_i; X_j, Y_j\right] &= \frac{1}{3}\sum_{k=1}^{K}\frac{1}{p_k}\left(-a_{ij\bullet} + a_{i\bullet\bullet} + a_{\bullet j\bullet} - a_{\bullet\bullet\bullet}\right)\left(I\left(Y_i = k\right) - p_k\right)\left(I\left(Y_j = k\right) - p_k\right) \\
&= \frac{1}{3}h_2\left(X_i, Y_i; X_j, Y_j\right),
\end{aligned} \tag{34}$$

where $a_{ij\bullet} = E[a_{ijl} \mid X_i, X_j]$, $a_{i\bullet\bullet} = E[a_{ijl} \mid X_i]$, $a_{\bullet j\bullet} = E[a_{ijl} \mid X_j]$, and $a_{\bullet\bullet\bullet} = E[a_{ijl}] = S_1$. Except for two extreme cases: (i) all $X_i$ are equal, and (ii) the categorial number $K = 1$, $Var[h_2(X_i, Y_i; X_j, Y_j)] > 0$. By the properties of V-statistic in Serfling (1980), we have:

$$\hat{V}_n = 3\frac{1}{n^2}\sum_{i,j=1}^{n} E\left[\frac{1}{3!}\sum\nolimits_3 b_{ijl} \mid X_i, Y_i; X_j, Y_j\right] + o_p\left(n^{-1}\right) = \frac{1}{n^2}\sum_{i,j=1}^{n} h_2\left(X_i, Y_i; X_j, Y_j\right) + o_p\left(n^{-1}\right). \tag{35}$$

Similar to the expression in Serfling (1980, Theorem 5.5.2), $h_2$ can be decomposed as

$$h_2(x_1, y_1; x_2, y_2) = \sum_{i=1}^{\infty} \lambda_i \phi_i(x_1, y_1) \phi_i(x_2, y_2),$$

where $E[\phi_i(X,Y)] = 0$ for all $i$, and

$$E[\phi_j(X,Y)\phi_k(X,Y)] = \begin{cases} 1, & j = k, \\ 0, & j \neq k. \end{cases}$$

Therefore, we have:

$$\begin{aligned}
\sum_{k=1}^{\infty} \lambda_k &= E\left[\sum_{k=1}^{\infty} \lambda_k \phi_k(X,Y)\phi_k(X,Y)\right] = E\left[h_2(X,Y;X,Y)\right] \\
&= \left(\frac{1}{c_p}\int_{\|\boldsymbol{\alpha}\|=1} E_{X_1}\left[Var_X\left(I\left(\boldsymbol{\alpha}^T X \leq \boldsymbol{\alpha}^T X_1\right)\right)\right] d\boldsymbol{\alpha}\right)\left(\sum_{k=1}^{K} \frac{1}{p_k} E\left[\left(I(Y=k) - p_k\right)^2\right]\right) \\
&= (S_1 + \pi S_3)(K-1).
\end{aligned}$$

Following the proof in Serfling (1980, Theorem 5.5.2), $\hat{V}_n$ has the following asymptotic property:

$$n\hat{V}_n \xrightarrow{D} \sum_{i=1}^{\infty} \lambda_i \chi_i^2, \tag{36}$$

where $\{\chi_i^2\}_{i=1}^{\infty}$ is a series of random variables following $\chi^2(1)$ distribution independently. According to the result in Eq.(31), $\widehat{LPCov_n}(X,Y)$ shows the same property. By Slutsky's Theorem,

$$n\widehat{LPC_n}(X,Y) \xrightarrow{D} \frac{1}{\Delta}\sum_{i=1}^{\infty} \lambda_i \chi_i^2.$$

The proof of Theorem 4 is complete.

□